\documentclass[aps,pre,twocolumn,groupedaddress]{revtex4-1}

\usepackage{xcolor}
\newcommand{\added}[1]{\textcolor{black}{#1}} 
\newcommand{\removed}[1]{\textcolor{red}{}} 
\usepackage[normalem]{ulem}

\usepackage{physics}
\usepackage{graphicx}
\graphicspath{ {./} } 

\begin{document}

\title{Effects of spherical confinement and backbone stiffness on flexible polymer jamming}


\author{Samuel M. Soik and Tristan A. Sharp}
\affiliation{Department of Physics and Astronomy, University of Pennsylvania, Philadelphia, Pennsylvania 19104, USA}


\date{\today}

\begin{abstract}
We use molecular simulations to study jamming of a crumpled bead-spring model polymer in a finite container and compare to jamming of repulsive spheres.
\added{After proper constraint counting, the onset of rigidity is seen to occur isostatically as in the case of repulsive spheres.
Despite this commonality, the presence of the curved container wall and polymer backbone bonds introduce new mechanical properties.
Notably, these include additional bands in the vibrational density of states that reflect the material structure as well as oscillations in local contact number and density near the wall but with lower amplitude for polymers.
\removed{An unexpected consequence is that}Polymers have fewer boundary contacts, and this low-density surface layer strongly reduces the global bulk modulus.
We further show that bulk-modulus dependence on backbone stiffness can be described by a model of stiffnesses in series and discuss potential experimental and biological applications.}
\removed{Polymer jamming occurs at lower density and with greater variability due to both the container wall and the polymer backbone.
In both cases, the onset of rigidity is marked by the creation of a single state of self-stress at isostaticity.
However, the confining wall significantly influences internal structure near the wall via density and contact number layering. Combined effects of layering and the backbone bonds are seen in the mechanical response, as reflected by the emergence of additional bands in the vibrational density of states, and in the bulk modulus, which is lower for polymers than monomers and varies with backbone stiffness according to a model of stiffnesses contributing in series.}

\end{abstract}


\maketitle

\section{Introduction \label{intro}}
																
The phenomenon of jamming is observed in systems ranging from granular materials flowing down a chute \cite{2dhopper} to \added{biomacromolecules \cite{proteinjamming,dnajamming1,dnajamming2,dnajamming3}.} \removed{DNA packed into a viral capsid.}
Most theoretical understanding of jamming comes from ideal models of granular materials, e.g., simulations of repulsive disks and spheres in periodic boundary conditions (PBCs).
In general, a packing of a biopolymer into a container involves, at least, two unavoidable additional features: backbone bonds that link the material into a polymer chain and container walls that \added{influence}\removed{change the structure of} the material near the boundary.
\removed{To investigate the consequences of these features, we extend the investigation of jamming to a material with unbreakable adhesive bonds in external spherical confinement (SC).}
\added{As a step toward closing the large gap between the most idealized models and experimental systems, we investigate the effects on jamming of unbreakable adhesive bonds and external spherical confinement (SC).}

Jamming occurs as the constituents of a flowing material sufficiently constrain one another's motion, leading to a configuration that resists applied stress.
A central question is whether the material jams \emph{isostatically}, that is, in precise balance of constraints and degrees of freedom, consistent with the boundary conditions.
Simulations of frictionless repulsive disks and spheres in PBCs have shown that the onset of rigidity occurs as a jump in particle-particle coordination number from zero to twice the dimensionality: four for disks and six for spheres, which correspond to isostaticity \cite{beforeepitome,epitome,jammingreview,finitescaling,granularpiles,phasediagram,silbertfriction,frictionshearjam}.
We show that jamming in our simulations with backbone bonds and a concave confining wall occurs isostatically in \added{fundamentally} the same way as for repulsive spheres.

Repulsive spheres typically jam at a packing fraction of about 64\%, which corresponds to the density of the maximally random jammed (MRJ) state \cite{rcp,mrj,packingofspheres,bernalcoordination,scottfollowup,
rcp,amorphousmetals,beforeepitome,epitome}. 
Unlike repulsive spheres, a bead-spring model polymer has ``built-in" constraints provided by backbone bonds.
Isostatic packings of \added{freely jointed}\removed{flexible} chains \added{of tangent hard spheres} can be obtained at $\phi^\mathrm{MRJ}$ using algorithms that eliminate the effects of connectivity and allow effective equilibration through \removed{entanglements and knotting}
\added{chain-connectivity-altering Monte Carlo moves \cite{mcscheme,fullflexphi,universalscaling,isopoly,
tangenthardspheres,xtalchains,xtalnucleation,softcolloidal,athermalpoly,mcconfinement,mcextreme}.} 
However, \removed{with more realistic dynamics}\added{when connectivity is preserved, approximately tangent fully flexible bead-spring} chains jam at about 2\% below $\phi^\mathrm{MRJ}$ in PBCs with little system-size dependence and retain a significant fraction of unconstrained degrees of freedom \cite{hoysemiflexible}.
Confinement \added{of monomers} also reduces the jamming density by inducing layering near the boundary 
\cite{containerwalls,vermanexp, 
quantumspheres, 
layering1, 
porosity,glassbeads, 
confinedgranular, 
layering, 
confinedrcp, 
spherespolyhedra, 
microcylinders}. 
We present both the reduction in density due to SC alone using repulsive sphere packings and the further reduction due to the polymer backbone \added{that links all particles together}.

On the other hand, few studies of polymer packings in confined geometries address mechanical properties.
Previous investigations have largely focused on chain conformation within the packing \cite{polysegmentdensity,polymermelts,metalchains,granularstiffpolymers,polystructure} 
and topological ordering of segments \cite{nikoubashmansemiflexible}.
Long polymers with specified bond-bond angles typically coil during packaging in SC to minimize bending energy \cite{ejectionforces,viraldna,bacteriophagep4,(semi)flex,dnatwist,dnaknotting,biopolymerreview} and thus exhibit boundary-induced layering \cite{dnatwist}.
In contrast, we use a crumpled flexible-chain model to avoid coiling \cite{(semi)flex} and to focus on the role of backbone connectivity in distinguishing the polymer from the monomer systems.

\added{
In Sec.~\ref{i&c}, we explain the necessity of using direct constraint counting rather than coordination number to assess the onset of rigidity due to unique considerations of systems in external confinement.}
\removed{We first explain the constraint-counting that causes jamming in the presence of external confinement and adhesive bonds.
We present the theory underlying our results and explain how, due to unique considerations of systems in external confinement, coordination number becomes less appropriate in assessing the onset of rigidity.}
In Sec.~\ref{sss}, we
show that \added{essentially} the same understanding of states of self-stress (SSSs) and zero modes in repulsive sphere packings \added{can be extended}\removed{extends} to the case of a polymer in SC.
In Sec.~\ref{ooj}, we provide the distribution of jamming densities in simulations of spherically confined polymers and compare to those of monomers in SC and in PBCs to isolate the effects of backbone bonds and the confining wall.
We find boundary-induced order in local density and coordination (Sec.~\ref{bis})
and in the vibrational density of states with effects on band structure due to the confining wall\removed{, pressure,} and the backbone  (Sec.~\ref{dos}).
\added{Finally, we show how the bulk modulus changes due to these structural differences between monomers and polymers as well as due to the polymer backbone stiffness (Sec.~\ref{bulkmodulus}).}
\removed{Finally, we discuss the effects of structural differences on the bulk modulus in comparing monomers to polymers 
as well as polymers with varying backbone stiffness (Sec.~\ref{bulkmodulus}).}

\section{Simulation details
\label{sd}}
To study jamming of flexible polymers, we use three-dimensional molecular dynamics simulations \cite{lammps} \added{of single chains, each composed of} $256\leq N \leq8192$ monodisperse frictionless spherical particles of diameter $\sigma$.
Each particle represents a monomer along a polymer chain, and interactions are governed by the following potentials: 
\begin{subequations}
\label{potentials}
\begin{align}
 V_0(r_{ij})&=\frac{\varepsilon_0}{2}\left(1-\frac{r_{ij}}{\sigma}\right)^2\theta\left(1-\frac{r_{ij}}{\sigma}\right),
 \\
 V_B(r_{kl})&=\frac{\varepsilon_B}{2}\left(1-\frac{r_{kl}}{\sigma}\right)^2, \\
 V_W(r_i)&=\frac{\varepsilon_W}{2}\left(\frac{1}{2}-\frac{R-r_i}{\sigma}\right)^2\theta\left(\frac{1}{2}-\frac{R-r_i}{\sigma}\right).
\end{align}
\end{subequations}
Nonconsecutive monomers interact via the harmonic repulsive potential $V_0(r_{ij})$, where $r_{ij}$ is the distance between the centers of particles (also referred to as \emph{sites}) $i$ and $j$, $\varepsilon_0$ is the characteristic energy, and $\theta(x)$ is the Heaviside step function.
Consecutive monomers $k$ and $l$ are bound by the two-sided harmonic potential $V_B(r_{kl})$ so that backbone bonds have energy scale $\varepsilon_B$ and rest length $\sigma$. 
To induce jamming, the polymer is confined by a spherical wall centered at the origin according to the radial harmonic potential $V_W(r_i)$, where $r_i$ is the radial coordinate of site $i$ and $R$ is the wall radius. 
The total potential energy $\mathcal{E}$ is the sum of all pairwise and wall potentials.
We consider at each site an equal point mass $m$, which sets the mass scale, and energies will be reported in units of $\varepsilon_0$, distances will reported in units of $\sigma$, pressures will be reported in units of $\varepsilon_0/\sigma^{3}$, and frequencies will be reported in units of $\sqrt{\varepsilon_0/m\sigma^2}$.

Disordered
configurations are generated by thermalizing the polymer chains at temperature $kT=0.003$ in a large confining sphere at packing fraction $\phi = N \left(\frac{\sigma}{2R}\right)^3 = 0.02$. 
Each thermal configuration is then quenched to $T=0$ using the FIRE algorithm \cite{fire}. We compress each quenched system in small increments of \added{$0.001\leq\Delta\phi\leq0.01$ (adjusted by system size)} by decreasing $R$ and minimizing energy after each compression until a jammed configuration is obtained, indicated by a nonzero $\mathcal{E}$.
We then expand or compress these configurations to within 1\% of each target pressure $p \equiv -\partial \mathcal{E}/\partial V$ where the system volume $V=4\pi R^3/3$ is that bounded by the confining sphere.
For each system of size $N$, at least 100 random configurations are prepared, and each of these is studied at a large range of target pressures $10^{-7}\leq p \leq10^{-1}$, bond energies $0.1\leq \varepsilon_B\leq 10$, and wall energies $0.1\leq \varepsilon_W\leq 10$.

The same procedures are repeated for \removed{unbonded}\added{nonbonded} monomers in SC (where $\varepsilon_B=0$) and in PBCs [where $\varepsilon_W=\varepsilon_B=0$ and $\phi = \frac{\pi N}{6} \left(\frac{\sigma}{2R}\right)^3$ in a cubic domain with side length $2R$].

\section{Isostaticity and coordination
\label{i&c}}

We review the analysis of the mechanical constraints that resist deformations and cause jamming.
This allows us to introduce the effects of confining walls and adhesive bonds.
\added{Here, we introduce the index theorem, and in Appendix~\ref{indexthm}, we derive the theorem in detail and explain associated subtleties.} \removed{((Section revised and moved to the appendix))}

\added{When interested in the linear response at low pressures, near jamming, we may consider the unstressed network of a given system by replacing all contacts (including backbone bonds and wall contacts) with unstretched harmonic springs in an analysis following Ref.~\cite{lubensky}.
The mapping to the spring system is exact in the limit of zero pressure, and each spring introduces one harmonic constraint.}
\added{
Each contact $i' \leq N_C$, where $N_C$ is the number of contacts, is replaced by a harmonic bond of rest length $r_{i'}$ equal to $r_{ij}$, $r_{kl}$, or $R-r_i$ [\added{referring to} Eqs.~(\ref{potentials})] depending on the interaction.
A zero mode is a normal mode of the system that causes no springs to be extended or compressed and corresponds to a motion with zero stiffness.
A SSS is a set of extensions and compressions assigned to the springs that results in zero net force at each site.
The index theorem embodies the fact that each contact either reduces the number of zero modes or increases the number of SSSs \cite{lubensky}, which, for a $d$-dimensional system with $dN$ degrees of freedom, is}
\begin{equation}
\label{eq:Index}
N_0-N_S=dN-N_C.
\end{equation}

Creating a rigid \added{(i.e., having no floppy modes)}, $d$-dimensional packing of spheres requires the number of constraints to match or exceed the degrees of freedom to be constrained \cite{maxwell}. Therefore, $N_C \geq dN-f(d)$, where $f(d)$ is the number of zero modes associated with rigid-body motions. 
PBCs permit $f(d)=d$ rigid translations whereas a frictionless $(d-1)$-spherical boundary permits $f(d) = \frac{1}{2}d(d-1)$ rigid rotations. 
By its strictest definition \cite{lubensky}, an isostatic system contains neither floppy modes nor SSSs [$N_0=f(d)$, $N_S=0$]; however, jammed packings necessarily have at least one SSS ($N_S\geq1$) corresponding to a nonzero modulus \cite{epitome,finitescaling} so that the number of contacts of a jammed isostatic system is
$N_C^\mathrm{iso}=dN-f(d)+1$.
Each additional constraint added to such a system creates an additional SSS,
\begin{equation}
\label{ns}
N_S=N_C-N_C^\mathrm{iso}+1.
\end{equation}

Constraints in repulsive sphere packings are commonly characterized by the average coordination number,
\begin{equation}
\label{coordnumber}
z=\frac{1}{N}\sum_{i=1}^N z_i,
\end{equation}
where $z_i$ is the number of contacts of particle $i$, but this is less appropriate in confinement.
First, without external confinement, as in PBCs, all contacts are between two particles, so $z = \frac{2N_C}{N}$ is twice the contact density, and the relation between $N_S$ and $z$ is
\begin{equation}
\frac{N_S}{N}=\frac{\Delta z}{2}\equiv\frac{z-z^\mathrm{iso}}{2},
\end{equation}
with $z^\mathrm{iso}=\frac{2N_C^\mathrm{iso}}{N}=2d-\frac{2f(d)}{N}$.
However, in external confinement, each wall contact involves only one particle.
Since the wall itself is not counted as a particle, wall contacts do not get double-counted, and the coordination number $z$ is lower than twice the contact density by an amount that decreases with system size,
\begin{equation}
\frac{2N_C}{N}-z=\frac{N_W}{N}\sim\frac{1}{L},
\end{equation}
where $N_W$ is the number of wall contacts and $L\equiv N^{1/d}$ is the linear system size.
Therefore, $z$ is twice the density of constraints only when each contact constrains two degrees of freedom.

Second, previous studies of monomers have removed rattlers in order to isolate the rigid subsystem \removed{for a clear computation of $\Delta z$}\added{so that $\Delta z$ is directly related to $N_S$} \cite{epitome,finitescaling}. Due to unbreakable bonds, polymers instead contain particles called flippers, which are constrained only by backbone bonds and thus can freely move tangent to their neighbors \added{\cite{fullflexphi}}. To accurately analyze the rigid subsystem of a confined polymer, \added{an analogous}\removed{a direct} computation of $\Delta z$ \added{would require} both a boundary correction and the removal of all flippers and the backbone bonds constraining them.

\section{Results}

\subsection{States of self-stress and zero modes
\label{sss}}

\begin{figure}
\includegraphics[scale = 0.4]{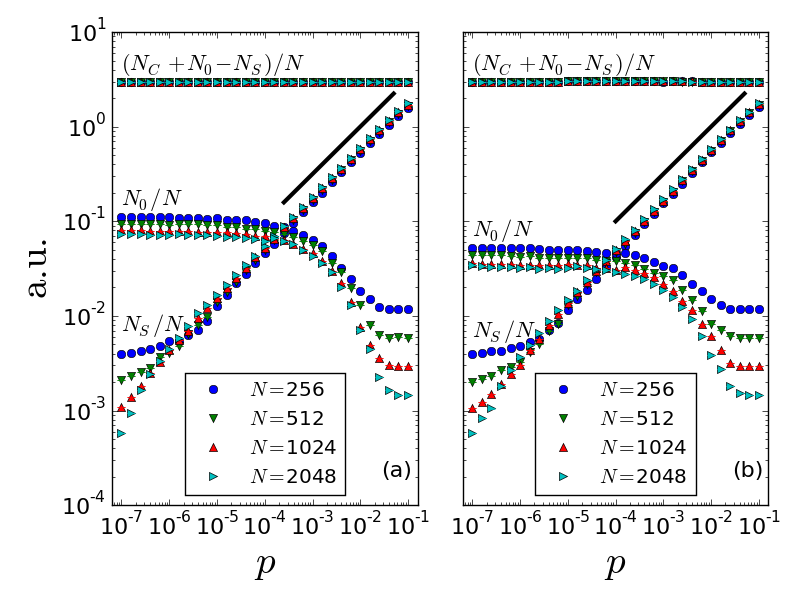}
\caption{Selected index theorem values for (a) monomers ($\varepsilon_B=0$, $\varepsilon_W=1$) and (b) polymers ($\varepsilon_B=\varepsilon_W=1$). Monotonically increasing (decreasing) curves show $N_S/N$ ($N_0/N$). Upper curves show computational results for $(N_C+N_0-N_S)/N$, equal to the dimensionality $d=3$ as guaranteed by Eq.~(\ref{eq:Index}). Black lines have slope $1/2$.
Approximately 100 states of each system size and pressure are considered. \label{index}}
\end{figure}

In Fig.~\ref{index}, we compute $N_S$ \added{[Eq.~(\ref{ns})]}.
Our results show that $N_S \geq 1$
as seen in the splitting of $N_S/N$ curves to $1/N$ in the low-$p$ limit.
Jamming in our systems, even with adhesive bonds and confinement, therefore corresponds to the introduction of a single SSS.
For the polymers, the SSSs may contain both extended and compressed backbone bonds; indeed, we find that $\approx 30\%$ of backbone bonds are extended near the jamming transition so the ratio of extended to compressed backbone bonds is $\approx 0.5$.

We see the power-scaling law $N_S/N \sim \Delta z \sim p^{1/2}$, the same as for spheres \cite{epitome,finitescaling} for both monomers and polymers ($\varepsilon_B=1$) in confinement.
This may be contrasted with a perfect $d$-dimensional crystal in external confinement, which would contain $N_S \gtrsim L^{d-1}$ at $p \to 0^+$.
The increasing number of SSSs involves an increasing number of sites ($N^\mathrm{rigid}$) and engaged contacts ($N_C^\mathrm{rigid}$) as the rigid subsystem grows.

To quantify the number of unconstrained motions, we compute $N_0$ \added{[Eq.~(\ref{n0})]}.
We find that $N_0^\mathrm{mono} >N_0^\mathrm{poly}$ in the low-$p$ limit.
For monomers, these are primarily rattlers, which have no constraints, so each contributes $d=3$ zero modes.
For polymers, these are primarily flippers;
the smaller number of zero modes reflects the extra constraints from the backbone bonds that constrain motion even on particles outside the rigid subsystem. \removed{
The two particles at the ends of the chain are called \emph{terminal} sites while all other particles within the chain are called \emph{internal}.
Terminal (internal) sites in polymer systems have at least 1 (2) constraint(s) due to backbone bonds; as such, terminal filaments contribute $(d-1)N_f$ 
zero modes, and internal filaments contribute
$(d-1)N_f-1$ zero modes.
(In the exceptionally rare case of two collinear, unstressed bonds around an internal site, 
only one constraint would be introduced, with two zero modes retained.) 
Here, \emph{filament} refers to a group of $N_f$ consecutive flippers on a polymer;
a filament is considered terminal if it contains a terminal site.
}\added{Because flippers can occur at chain ends and may involve consecutive polymer sites, directly}\removed{Directly} computing the precise number of flippers \added{from $N_0$} requires distinguishing topologically distinct \added{groups}\removed{filament types}
and is not necessary to see that about 1 to 2\% of the degrees of freedom are unconstrained even at moderate pressures. The significant number of unconstrained motions is consistent with other realistic packing protocols \cite{nonmonotonic,polymelts}.

The fraction of rattlers (flippers) decreases with system size. In the high-$p$ limit, no rattlers (flippers) remain as all particles become sufficiently coordinated that the only remaining zero modes are those associated with $f(d)$ rigid rotations within the spherical container.
\added{As pressure increases, particles rearrange to allow the system to relax. Rearrangements only result in small-scale configurational changes, even though the chain spans the full system.}

Next, we delete rattlers and flippers, isolating the $N^\mathrm{rigid}$ particles and $N_C^\mathrm{rigid}$ engaged contacts of the rigid subsystem.  At all pressures, we find that the number of zero modes that remain is again $f(d)$, indicating that no other zero modes are present in the rigid subsystem. Therefore, from Eq.~(\ref{eq:Index}),
\begin{equation}
\lim_{p \to 0^+} N_C^\mathrm{rigid} = dN^\mathrm{rigid} - f(d) + 1 = N_C^\mathrm{rigid,iso},
\end{equation}
and we find that the rigid subsystem jams isostatically.

\subsection{Packing fraction at jamming
\label{ooj}}

\begin{figure}
\includegraphics[scale = 0.5]{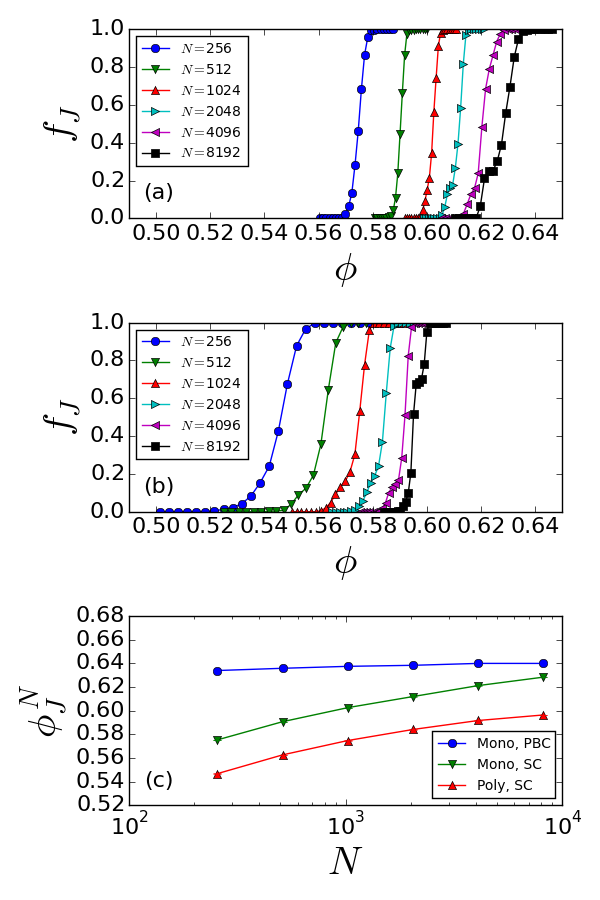}
\caption{Fraction of jammed states for (a) monomers and (b) polymers ($\varepsilon_B=1$) in SC ($\varepsilon_W=1$). (c) Comparison of $\phi_J^{N,\mathrm{mono}}$ and $\phi_J^{N,\mathrm{poly}}$ in SC to monomers in PBCs. Approximately 500 states (250 for the largest systems) of each system size are considered.
\label{jdist}}
\end{figure}

\added{For reference, we provide}\removed{Fig.~\ref{jdist} shows} the fraction of systems that are jammed $f_J$ at packing fraction $\phi$ as well as the average packing fraction at jamming $\phi_J^N$ for $256 \leq N \leq 8192$ \added{(Fig.~\ref{jdist})}.

Monomers in PBCs jam near 64\% as expected for MRJ states for all system sizes.
Confinement shifts jamming distributions to lower densities and increases system-size dependence.
$\phi_J^{N,\mathrm{mono}}<\phi^\mathrm{MRJ}$, in agreement with previous studies of confined monomers \cite{porosity,glassbeads,confinedrcp,spherespolyhedra}. 
The deviation of $\phi_J$ from $\phi^\mathrm{MRJ}$ is almost 6\% at $N=256$ and diminishes to less than 1\% by $N=8192$.

Figures~\ref{jdist}(b) and~\ref{jdist}(c) show that the inclusion of unbreakable backbone bonds further reduces the jamming density to almost 10\% below $\phi^\mathrm{MRJ}$ at $N=256$ and 4\% below at $N=8192$.
\removed{We see that $\phi_J^{N,\mathrm{mono}}>\phi_J^{N,\mathrm{poly}}$ in SC, due to the constraints arising from the polymer backbone.}
\removed{Backbone bonds prevent consecutive polymer sites from drifting apart during the approach to jamming and also lead to attractive forces in extension.
Constraining effects of the backbone do not strongly diminish with system size; the
}\added{The} $\approx 4\%$ difference between $\phi_J^{N,\mathrm{mono}}$ and $\phi_J^{N,\mathrm{poly}}$ persists across system sizes, similar to the density shift seen in jamming of flexible thermal polymers in PBCs \cite{hoysemiflexible}.
Backbone bond stiffness has no appreciable effect on $\phi_J^{N,\mathrm{poly}}$ of flexible polymers, so only $\varepsilon_B=1$ data are shown in Fig.~\ref{jdist}.

\added{
In addition, we consider monomer packings generated from jammed polymer configurations by deleting the backbone bonds. Without the extended bonds, the packings are unstable, and jamming is reattained at densities similar to the monomer distributions in Fig.~\ref{jdist}(a).
}

\removed{Finally, the width of the jamming distribution substantially widens with decreasing system size, especially for polymers.}

\subsection{Boundary-induced structure
\label{bis}}

\begin{figure}
\includegraphics[scale = 0.4]{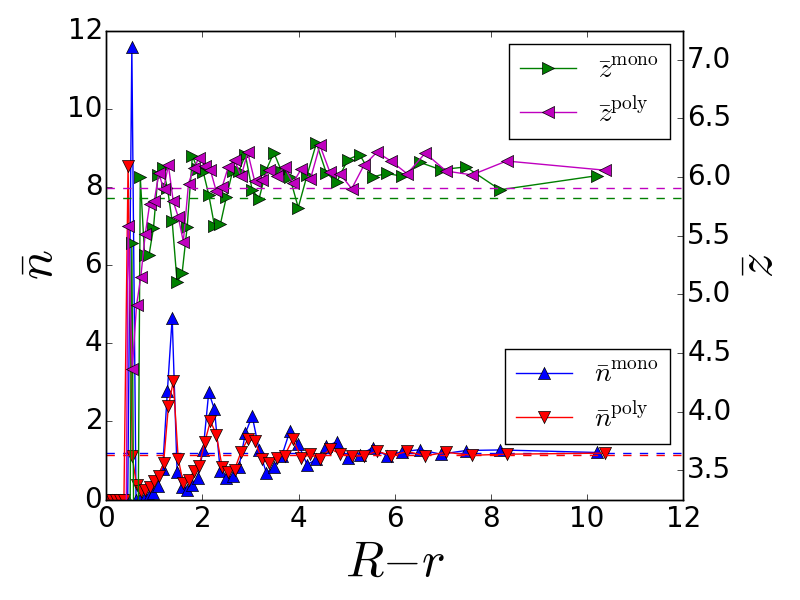}
\caption{Average local number density $\bar{n}$ (lower curves, left axis) and average local coordination $\bar{z}$ (upper curves, right axis) for $N=8192$ systems ($\varepsilon_W=\varepsilon_B=1$) at $p=10^{-4}$.
Dashed horizontal lines are the global number density $n$ and coordination number $z$ for each system. \added{Some 50 bins of equal volume were used.} \removed{Curves are averages of approximately 100 configurations.}
\label{nz}
}
\end{figure}

We compute the average local number density $\bar{n}$ and average local coordination $\bar{z}$ by binning point masses at $\{\textbf{r}_i\}$ and their respective coordination values $\{z_i\}$ over distance from the boundary $R-r$ (Fig.~\ref{nz}). \added{Here, each nonbonded contact, backbone bond, and wall contact involving particle $i$ is included as one contact in $z_i$.} Density layering is significant near the boundary (and, as reflected in Fig.~\ref{jdist}, reduces $\phi^N_J$).
The global number density $n$ and coordination number $z$ are shown as dashed lines in Fig.~\ref{nz}.
Oscillations occur in both $\bar{n}$ and $\bar{z}$, similar to previous density profiles of confined monomers determined in experiments \cite{glassbeads,microcylinders} and simulations \cite{quantumspheres,spherespolyhedra,layering1,confinedgranular,layering,confinedrcp,dnatwist} \added{as well as tangent hard-sphere chains \cite{mcconfinement}}.
Both oscillatory periods are consistent with the height of a regular tetrahedron (3-simplex) $\sqrt{2/3}\sigma \approx 0.82\sigma$ and agree with the well-established polytetrahedral structure of jammed monomer \cite{icosa-tetra-hedra,polytetrahedra} and polymer \cite{tangenthardspheres,xtalchains,xtalnucleation,softcolloidal,athermalpoly,mcconfinement,hoysemiflexible,mcextreme} states.
We note that sharply peaked maxima (minima) in $\bar{n}$ ($\bar{z}$) are separated by broad rounded minima (maxima).
This qualitative ``inversion" of curves would suggest that sites of high-density layers are, perhaps unintuitively, \emph{less} coordinated than sites in the low-density layers between them.
This could be rationalized by considering that particles in high-density layers sit between two lower-density layers with which they have fewer contacts than particles in low-density layers that sit between two high-density layers.
However, the structure is even more complex than this as $\bar{z}$ curves are also shifted to the right of their inverted $\bar{n}$ counterparts; qualitatively, this phase shift appears to be about one-quarter of the period.

Although the curves are similar for monomers and polymers, a first noticeable difference is the height of the initial narrow peak at $R-r=\sigma/2$,
indicating $N_W^\mathrm{mono} > N_W^\mathrm{poly}$, which becomes important to the bulk modulus as considered in Sec.~\ref{backboneconnectivity}. Additionally, for polymers, the oscillation amplitude of $\bar{n}$ is noticeably less than that of monomers, indicating that polymers exhibit less-extreme layering.
In contrast, $\bar{z}^\mathrm{mono}<\bar{z}^\mathrm{poly}$ at nearly all points because of backbone bonds retained by flippers, which lead to the higher global coordination $z$ of polymers than monomer systems with fully uncoordinated rattlers.

\subsection{Density of states
\label{dos}}

To investigate the vibrational density of states, we construct the \removed{$dN \times dN$}dynamical matrix $D_{i\nu}^{j\mu}$ \added{[Eq.~(\ref{dm})]}.\removed{((Section moved to the Appendix.))}
The set of eigenvectors $\{U_i^\mu\}$ of \added{$D_{i\nu}^{j\mu}$}\removed{the matrix} are the polarization vectors of the system's normal modes, and the eigenvalues $\{\lambda\}=\{\omega^2\}$ are the squared frequencies of the normal modes \cite{ashcroft}.
From $\{\omega\}$, we compute the density of states $D(\omega)$.
Since there is little variation among system sizes, we present only $N=2048$ data.

\subsubsection{Boundary modes
\label{boundary}}

\begin{figure}
\includegraphics[scale = 0.4]{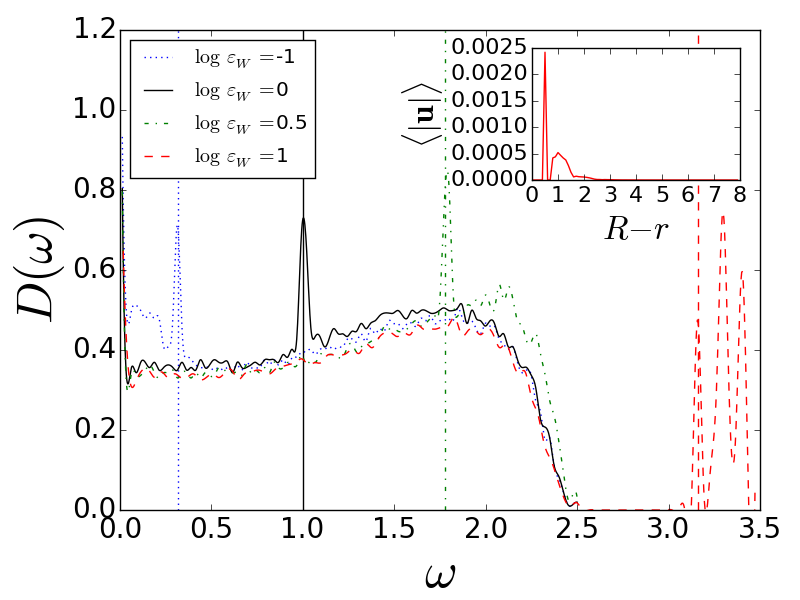}
\caption{$D(\omega)$ at $p=10^{-4}$ and $0.1 \leq \varepsilon_W \leq 10$ for a $N=2048$ monomer system. Vertical lines indicate corresponding $\omega_W$ values. The inset: average displacement for eigenstates $\{U_i^\mu:\omega>3,\varepsilon_W=10\}$. \label{boundarymodes}}
\end{figure}

We compute \removed{the density of states}$D(\omega)$ in systems with wall potentials $0.1\leq\varepsilon_W\leq10$ for monomers in SC (Fig.~\ref{boundarymodes}). 
We first note that peaks at $\omega=0$ represent zero modes due to rattlers and rigid rotations. 
The curves have the universal characteristic shapes seen previously in disordered systems in PBCs \cite{epitome}, the so-called boson peak at small finite $\omega$.
However, wall potentials induce $N_W$ boundary modes with typical frequencies of $\omega_W \equiv \sqrt{\varepsilon_W/m\sigma^2}$, resulting in additional pronounced peaks.

At large $\varepsilon_W$, the additional modes lead to a band gap in $D(\omega)$.
In this case, modes with $\omega>3$ may be isolated, and we bin the total set of polarization magnitudes $\{|\textbf{u}_i|\}$ over $R-r$ to compute the average polarization $\langle|\textbf{u}|\rangle$ with respect to distance from the wall (Fig.~\ref{boundarymodes}, the inset).
The boundary modes are almost entirely localized to the two layers of sites nearest the boundary, giving the two distinct peaks in $\langle|\mathbf{u}|\rangle$.

\removed{((Removed section: pressure effects on D.O.S.))}

\subsubsection{Backbone modes
\label{backbone}}

\begin{figure}
\includegraphics[scale = 0.5]{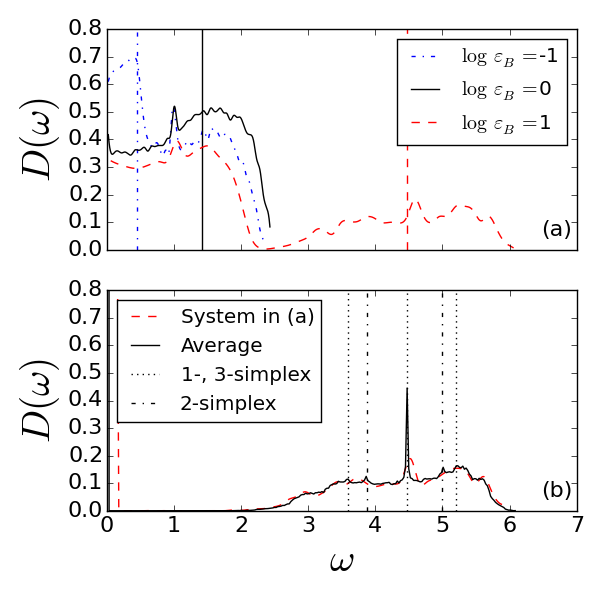}
\caption{$D(\omega)$ for $N=2048$ polymer systems at $p=10^{-4}$ and $0.1\leq\varepsilon_B\leq10$. (a) Three systems with $\varepsilon_0=\varepsilon_W=1$. Vertical lines indicate $\sqrt{2}\omega_B$. 
(b) $D(\omega)$ with $\varepsilon_0=\varepsilon_W=0$, $\varepsilon_B=10$ for the system in (a) and averaged over $>20$ systems. Vertical lines indicate natural frequencies of regular simplices.  
\label{bondmodes}}
\end{figure}

\removed{Finally, we vary}\added{Next, we see the effect of backbone-bond stiffness on the density of states} [Fig.~\ref{bondmodes}(a)]. 
Backbone interactions lead to a broad band \removed{of vibrational states,} approximately centered at $\sqrt{2}\omega_B\equiv\sqrt{2\varepsilon_B/m\sigma^2}$ as identified in Refs.~\cite{polymerdos,
vdos}.
The broadness of this band may be contrasted with the narrower and more structured boundary-mode band in $D(\omega)$.
Like the high-$\varepsilon_W$ boundary band in Fig.~\ref{boundarymodes}, the high-$\varepsilon_B$ backbone band's separation from the bulk band suggests a degree of independence in mode structure, and the density of states of the full system can be broken down into contributions from all three sources. 

In Fig.~\ref{bondmodes}(b), we replot $D(\omega)$ when $\varepsilon_B=10$ for the system in Fig.~\ref{bondmodes}(a) but set $\varepsilon_0=\varepsilon_W=0$ in our computation of $K_{i'}^{j'}$ (see Appendix~\ref{indexthm}).
Bulk and boundary bands vanish into the $\delta$-function peak of zero modes, but we observe almost no change in the backbone band,
highlighting its independence from the bulk band.
A universal feature of the polymer vibrational spectra, the broad backbone band is a feature of real globular proteins \cite{globular1,globular2}.
For better resolution of its features, we compute the average curve from $>20$ systems.
Several pronounced peaks appear in the backbone band, which are similar to the signatures of analytically derived modes in collections of short chains of length $N_\mathrm{ch}\leq5$ in PBCs \cite{vdos}.
The most pronounced peak is at $\omega=\sqrt{2}\omega_B$, which corresponds to the vibrational frequency of the 1-simplex (a single bond) as well as a normal mode of the general 3-simplex.
There are also small peaks at $\omega=\sqrt{2\pm\frac{1}{2}}\omega_B$ and $\omega=\sqrt{2\pm\sqrt{\frac{1}{2}}}\omega_B$, which correspond to vibrational frequencies of regular 2- and 3-simplices, respectively.

\subsection{Bulk modulus
\label{bulkmodulus}}
\subsubsection{Effect of backbone connectivity
\label{backboneconnectivity}}

\begin{figure}
\includegraphics[scale = 0.4]{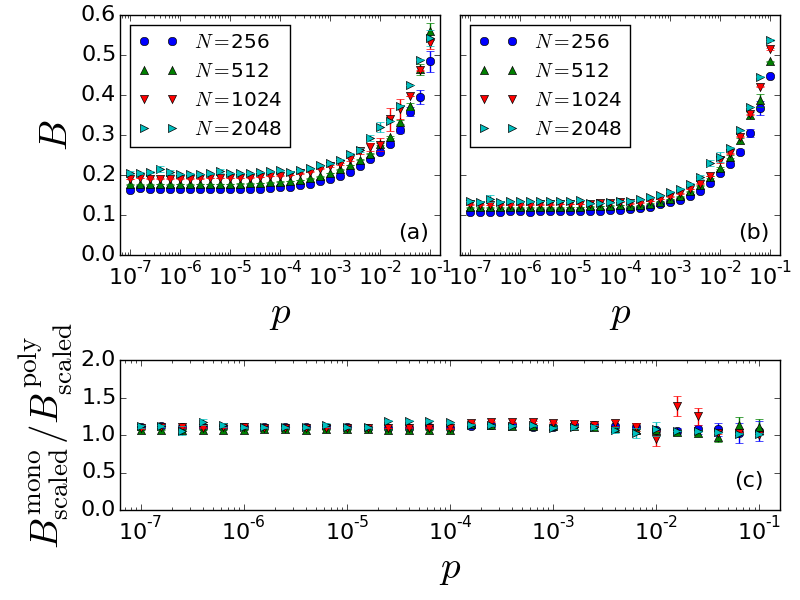}
\caption{Bulk modulus for (a) monomer and (b) polymer systems, (c) \removed{scaled $B$}\added{$B_\mathrm{scaled}^\mathrm{mono}/B_\mathrm{scaled}^\mathrm{poly}$} ratio.
\label{bcomp}}
\end{figure}

Plotting the bulk modulus $B\equiv \phi \partial p/\partial\phi$ of monomers and polymers over a range of $10^{-7}\leq p \leq10^{-1}$ (Fig.~\ref{bcomp}), we find a constant, nonzero limit $\lim_{p\to0^+}=B_0$, consistent with the power-law scaling relation $B\sim p^0$ \cite{epitome,jammingreview}.
As pressure increases from zero, $B$ remains within 1\% of $B_0$ until $p \sim 10^{-4}$
whereas over this range $N_S$ increases by orders of magnitude from $N_S=1$ in the system sizes considered here (Fig.~\ref{index}).
$B$ also varies with $N$, mostly due to variation of $\phi_J^N$ with system size (Fig.~\ref{jdist}).

The bulk modulus is substantially ($\approx40$\%) higher for monomers than for polymers.
Variation in the prefactor $\phi$ in the definition of $B$ accounts for only a small part of the difference; $\phi_J^{N,\mathrm{poly}}$ is only $\approx4$\% lower than $\phi_J^{N,\mathrm{mono}}$ [Fig.~\ref{jdist}(c)].
Therefore, it must also be that $(\partial p/\partial\phi)^{N,\mathrm{mono}}>(\partial p/\partial\phi)^{N,\mathrm{poly}}$. 
Section~\ref{sss} showed that the rigid subsystems are nearly equal in size between the two system types ($N^\mathrm{rigid,mono} \approx N^\mathrm{rigid,poly}$, $N_C^\mathrm{rigid,mono} \approx N_C^\mathrm{rigid,poly}$),
so the difference in $B$ must be due to configurational differences.

Recall that monomer packings have stronger layering and far more wall contacts than polymers (Sec.~\ref{bis}).
Only wall contacts couple the motion of the wall to the interior packing, and therefore we may expect $B$ to rise with the wall contact density $N_W/A \sim N_W(\phi/N)^{2/3}$.
\added{We consider the modulus scaled correspondingly $B_\mathrm{scaled} \equiv \frac{B}{N_W} \left( \frac{N}{\phi} \right)^{2/3}$ and plot the ratio $B_\mathrm{scaled}^\mathrm{mono}/B_\mathrm{scaled}^\mathrm{poly}$} in Fig.~\ref{bcomp}(c).\removed{we plot the ratio $\frac{B^\mathrm{mono}N_W^\mathrm{poly}}{B^\mathrm{poly}N_W^\mathrm{mono}}\left(\frac{\phi^\mathrm{poly}}{\phi^\mathrm{mono}}\right)^{2/3}$.} We see that this ratio is approximately $1$ for all system sizes and pressures, demonstrating that the difference in $B$ is primarily due to $N_W$.

\subsubsection{Effect of backbone stiffness
\label{backbonestiffness}}

\begin{figure}
\includegraphics[scale = 0.4]{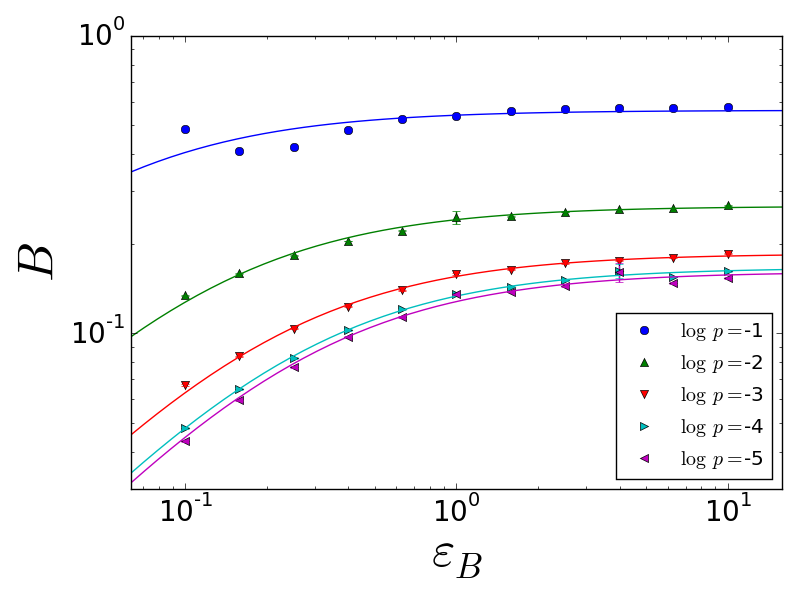}
\caption{Bulk modulus for $N=2048$ polymer systems. Solid lines show curve fitting to Eq.~(\ref{beq}). 
\label{bbond}}
\end{figure}

\begin{table}
\caption{Curve-fitted parameters for Eq.~(\ref{beq}). \label{btable}}
\begin{ruledtabular}
\begin{tabular}{ c c c }
$\mathrm{log}\, p$ & $B_\infty$ & $\epsilon$ \\
\hline
-1 & 0.562 $\pm$ 0.005 & 0.039 $\pm$ 0.004 \\
-2 & 0.2676 $\pm$ 0.0006 & 0.110 $\pm$ 0.002 \\
-3 & 0.1854 $\pm$ 0.0004 & 0.194 $\pm$ 0.002 \\
-4 & 0.1663 $\pm$ 0.0004 & 0.247 $\pm$ 0.003 \\
-5 & 0.1612 $\pm$ 0.0008 & 0.260 $\pm$ 0.006 \\
\end{tabular}
\end{ruledtabular}
\end{table}

We also investigate the effect on $B$ of the backbone stiffness by varying $\varepsilon_B$ into both low-stiffness and high-stiffness regimes at pressures $10^{-5} \leq p \leq 10^{-1}$, plotted in Fig.~\ref{bbond}.
At low pressures, the bulk modulus vanishes if $\varepsilon_B \to 0$ as the configuration without backbone bonds is undercoordinated for rigidity. The bulk modulus saturates to a constant as $\varepsilon_B \to \infty$;
backbone bonds become essentially inextensible compared to other contacts, yet the material can still deform around an infinitely stiff backbone.
(In the equivalent case of decreasing $\varepsilon_0$, recall that the units of $\varepsilon_B$ and $B$ are proportional to $\varepsilon_0$ so that $B$ decreases proportionally to $\varepsilon_0$.)

To motivate a simple curve-fitting relation, consider that the material is isostatic at jamming, so the existence of the bulk modulus is dependent on every contact, similar to the simple situation of springs all in series.
Given that $B$ is measured by isotropically deforming the wall, we therefore consider a different system: a one-dimensional chain of $N_0^\mathrm{eff}$ springs of stiffness $k_0\equiv\varepsilon_0/\sigma^2$ (these represent both wall and nonbonded-particle interactions since we have set $\varepsilon_0=\varepsilon_W=1$) and $N_B^\mathrm{eff}$ springs of stiffness $k_B\equiv\varepsilon_B/\sigma^2$ (representing backbone interactions).
The chain's overall effective spring constant is
\begin{equation}
k_\mathrm{eff} = \left(\frac{N_0^\mathrm{eff}}{k_0}+\frac{N_B^\mathrm{eff}}{k_B}\right)^{-1},
\label{keff}
\end{equation}
which is proportional to the bulk modulus $B = \beta k_\mathrm{eff} / \sigma$, where $\beta$ is a dimensionless constant.
Rearranging Eq.~(\ref{keff}) in terms of $\varepsilon_B$ and $B_\infty=\lim_{\varepsilon_B \to \infty}B$ yields
\begin{equation}
B=B_\infty (1+\epsilon/\varepsilon_B)^{-1},
\label{beq}
\end{equation}
with $B_\infty=\beta k_0/\sigma N_0^\mathrm{eff}$ and $\epsilon=\sigma^2k_0N_B^\mathrm{eff}/N_0^\mathrm{eff}$.
In natural units $\sigma=k_0=1$, the energy scale $\epsilon$ represents the ratio $N_B^\mathrm{eff}/N_0^\mathrm{eff}$.

We plot curve fits using Eq.~(\ref{beq}) in Fig.~\ref{bbond}, which agree well with data for $p\leq10^{-2}$; curve-fitted values of $\epsilon$ and $B_\infty$ are given in Table~\ref{btable}. The upward deviation in our data at $p=10^{-1}$ for the lowest $\varepsilon_B$ is a result of extreme compression and overcoordination as second-nearest-neighbor interactions occur, which the fitting form is not meant to capture. Pressure effects diminish in the low-$p$ limit.

\section{Discussion and conclusions}

We have analyzed jammed configurations of a flexible bead-spring polymer in SC. 
Despite the presence of adhesive backbone bonds and spherical confining walls, the conditions at jamming \added{superficially}\removed{largely} carry over from the case of repulsive spheres in PBCs. After accounting for the rigid-body motions within the spherical container, wall contacts, and underconstrained particles (rattlers and flippers), we see that jamming occurs exactly at isostaticity and coincides with the emergence of a single SSS.
Jamming occurs at somewhat reduced density compared to monomers, and, upon further compression, the number of SSSs scales as the square root of pressure as for monomers in SC.

\removed{Confining walls and the polymer backbone influence the internal structure, density of states, and bulk modulus of the jammed material.}
The boundary causes layering in both local density and coordination, which are, unexpectedly, out of phase; qualitatively, curves for density and coordination are inverted in shape and phase-shifted $\approx\pi/2$.
The boundary also introduces a narrow band of vibrational modes into the density of states with characteristic frequency scaling with the square root of the wall stiffness.
At high wall stiffness, these modes are highly localized to the outermost two layers of sites.
The independence of boundary modes from bulk modes extends to backbone modes; bands generated by high-stiffness backbone bonds are virtually unchanged after the removal of nonbonded and boundary potentials.
Not only do these bands follow a universal pattern, but they also display peaks corresponding to regular low-dimensional simplices,
indicating the possibility of inferring aspects of the internal structure from the vibrational spectrum.

The higher number of wall contacts in monomer packings raises the bulk modulus by $\approx40\%$ compared to polymers.
An explanation comes from a model of stiffnesses in series that scales with the wall contact number.
A similar conceptual model motivates a fitting relation that describes the dependence of the bulk modulus on backbone stiffness \added{and predicts its value in the limiting case of incompressible backbone bonds}.

\added{Although packing of a flexible-chain polymer is a highly idealized model of a biopolymer, several insights may apply immediately to experiment.
The vibrational states convey information about the strength of confinement, the number of boundary constraints, and the backbone configuration, which could be exploited to study and potentially manipulate polymer structure.
Our results may also apply to the cytoskeleton, the protein network that spans the cell from the nucleus to the cell membrane and accounts for cytoplasmic structure and rigidity.
The number of contact points with the cell membrane may be strongly linked with cellular compressibility and membrane flexibility.
This dependence could be measured experimentally, e.g., via atomic force microscopy.
Further biological relevance could be found within the cell nucleus where our model may help elucidate the envelope's influence on chromatin structure and mechanics.
In addition, we hope this paper clarifies fundamental aspects of jamming with regard to internal constraints and the finite boundaries present in all real systems.}

Future analysis may investigate the \added{spatial structure of SSSs in SC, the origin of the apparent phase shift in local density and coordination, or the} material elasticity at higher pressure and with higher-curvature walls where internal stresses and higher-order terms in the energy expansion are relevant \added{(discussed in Appendix~\ref{constraintcounting})}.
A fuller analysis may also \removed{\added{probe}\removed{investigate} the spatial structure of SSSs in SC or}consider \removed{the effects of variable backbone stiffness, backbone-bending and dihedral stiffness, bond stresses, or attractive self-interactions.}\added{nonbonded adhesive interactions, backbone-bending stiffness, dihedral stiffness, bond stresses, or finite temperature to yield more accurate models of real biopolymers.}

\begin{acknowledgments}
We are grateful for helpful discussions with A. Liu, C. Goodrich, and T. Lubensky. This research was funded, in part, by the U.S. DOE, Office of Basic Energy Sciences, Division of Materials Sciences and Engineering (Award No. DE-FG02-05ER46199). Computational support was provided by the LRSM HPC cluster at the University of Pennsylvania.
\end{acknowledgments}


\appendix

\section{Unstressed-network approximation
\label{constraintcounting}}

\added{We first explain the approximation of low stresses, accurate at low pressures, that permits the index theorem analysis based on an unstressed spring network (Appendix~\ref{indexthm}).
In the following, we use Einstein notation, and sites are labeled by plain Roman indices, bonds are labeled by primed Roman indices, and Cartesian components are labeled by Greek indices.}

We explicitly calculate the lowest-order energy terms to analyze stability of a static configuration $\{\textbf{r}_i^0\}$ after energy minimization, i.e., in force balance, with potential energy $\mathcal{E}_0 = \mathcal{E}(\{\textbf{r}_i^0\})$.
Let $\textbf{r}_i$ be the position of particle $i$ and $\textbf{u}_i = \textbf{r}_i - \textbf{r}_i^0$ be its displacement from its reference position. 
For small displacements, we can Taylor expand the energy,
\begin{equation}
\mathcal{E}(\{\textbf{r}_i\}) = \mathcal{E}_0 + u_i^\mu \frac{\partial \mathcal{E}}{\partial r_i^\mu} \bigg|_{\{\textbf{r}_i^0\}}  + \\
\frac{u_i^\mu u_j^\nu}{2} \frac{\partial^2 \mathcal{E}}{\partial r_i^\mu \partial r_j^\nu} \bigg|_{\{\textbf{r}_i^0\}} + \cdots ,
\end{equation}
\added{where terms proportional to $u_i^\mu$ are zero since we expand about a stable configuration.} 
\removed{The contribution proportional to displacement is necessarily zero since we expand about a minimum where the force on each particle vanishes ($-\partial \mathcal{E}/\partial r_i^\mu = 0$).}

\begin{figure}
\includegraphics[scale = 0.6]{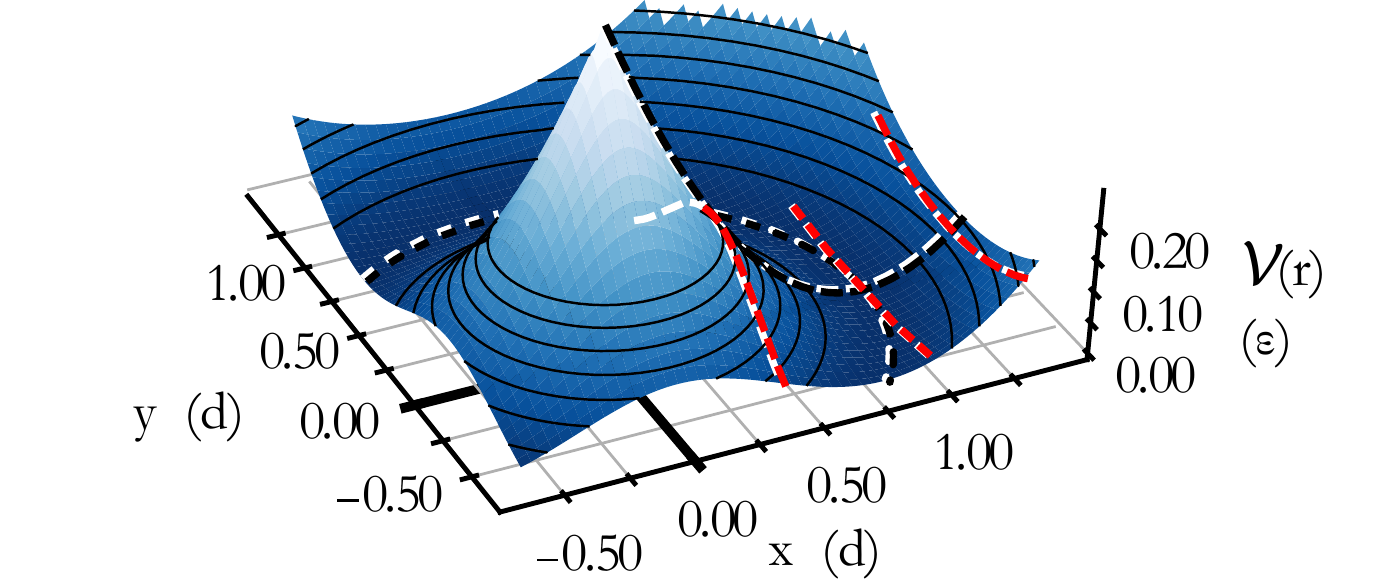}
\caption{Plot of the $z=0$ potential energy surface $\mathcal{V}(r)=\varepsilon(1-r/d)^2/2$, $r=\sqrt{x^2+y^2+z^2}$ to illustrate the curvature in the tangential direction of an extended or compressed harmonic interaction and the higher-order stabilizing terms at zero pressure, which are absent in the case of repulsive interactions.
\label{mexicanhat}}
\end{figure}

All potentials in the simulation where nonzero have the form $\mathcal{V}(r)=\varepsilon(1-r/d)^2/2$, \added{where $r=|\mathbf{r}|$ corresponds to displacements $|\mathbf{r}_i-\mathbf{r}_j|$, $|\mathbf{r}_k-\mathbf{r}_l|$, or $|\mathbf{r}_i|$ [referring to Eqs.~(\ref{potentials})].}
A displacement component parallel to the interaction direction $u_{\parallel} \equiv \textbf{u}  \cdot \textbf{r}/r$ corresponds to stiffness 
$\kappa \equiv 
\partial^2 \mathcal{V}/\partial u_{\parallel}^2 = \varepsilon/d^2$.
A component perpendicular $u_{\perp} \equiv |\textbf{u} - u_{\parallel} \textbf{r}/r|$ also has finite stiffness $\partial^2 \mathcal{V}/\partial u_{\perp}^2 = \kappa (1-d/r)$, which is positive for wall contacts and extended backbone bonds.
Explicitly, the change in energy due to a small displacement perpendicular to $\mathbf{r}_{ij}$, $\mathbf{r}_{kl}$, or $\mathbf{r}_i$ is
\begin{subequations}
\label{expansions}
\begin{align}
 \Delta V_0(u_\perp) &= \frac{\varepsilon_0}{2}\left(1-\frac{\sigma}{r_{ij}}\right) \left(\frac{u_\perp}{\sigma}\right)^2 + O(u_\perp^4), \\
 \Delta V_B(u_\perp) &= \frac{\varepsilon_B}{2}\left(1-\frac{\sigma}{r_{kl}}\right) \left(\frac{u_\perp}{\sigma}\right)^2 + O(u_\perp^4), \\
 \Delta V_W(u_\perp) &= \frac{\varepsilon_W}{2}\left(1-\frac{R-\sigma/2}{r_i}\right)\left(\frac{u_\perp}{\sigma}\right)^2 + O(u_\perp^4),
\end{align}
\end{subequations}
where $\theta(x)$ is omitted for brevity.
The prefactor of the quadratic term is negative for overlapping monomers ($r_{ij}<\sigma$), so energy decreases in the perpendicular direction, and the particles tend to slip off one another.  Small displacements perpendicular to extended backbone bonds ($r_{kl}>\sigma$) or tangential to the wall instead require an increase in energy, resulting in linear restoring forces. 
Prefactors vanish in the unstressed, i.e., zero-energy limit ($r=d$), so the energy costs of these motions appear only at $O(u_\perp^4)$ and produce no linear response.
\added{Figure~\ref{mexicanhat} illustrates the tangential curvature of $\mathcal{V}(r)$ for $r<d$ ($\partial^2 \mathcal{V}/\partial u^2_\perp<0$), $r=d$ ($\partial^2 \mathcal{V}/\partial u^2_\perp=0$), and $r>d$ ($\partial^2 \mathcal{V}/\partial u^2_\perp>0$).}

\added{Although extended backbone bonds and wall contacts constrain tangential motion,} we focus on the unstressed case (at jamming) where the mapping to unstretched springs is exact for harmonic analysis \cite{lubensky}.
In that case, each interaction \added{constrains motion only along the interaction direction.}
Therefore, at sufficiently low pressure, only relative motion in the direction normal to the contact contributes significantly to the linear response, but higher-order terms in the energy expansion can, in principle, affect jamming in some materials, e.g., they are seen to stabilize zero-frequency modes in packings of aspherical particles \cite{hypostaticity,ellipses,ShattuckOHern2018}. 
At rest length, nonzero contributions up to fourth order in the expansion come from the terms $\kappa u_{\parallel} u_{\perp}^2 / 6d$, $-\kappa u_{\parallel}^2 u_{\perp}^2 / 12d^2$, and $\kappa u_{\perp}^4 / 8 d^2$, so
the curvature of confining walls and the adhesive regime of backbone bonds may contribute to higher-order stability at zero pressure. 
In principle, these terms may be able to stabilize zero modes in the packing; 
however, as stated in the main text, after rattlers and flippers have been deleted, no zero-frequency modes are present in our packings other than rigid-body motions, indicating that harmonic analysis accounts for all constraints in our sphere packings.

\section{Derivation of the index theorem
\label{indexthm}}

Site displacements form the $dN$-dimensional displacement vector $U_i^\mu$, where $\mu$ indexes the $d=3$ Cartesian components of each vector $\textbf{u}_i$.
The linear operator $C_{i'\mu}^i$, termed the compatibility matrix, maps $U_i^\mu$ to the $N_C$-dimensional bond elongation vector $E_{i'} \equiv \frac{\partial r_{i'}}{\partial r_i^\mu} U_i^\mu$,
\begin{equation}
\label{compatibilitymatrix} 
C_{i'\mu}^i U_i^\mu=E_{i'} .
\end{equation}
Since a zero mode is described by a set of displacements that causes no bond elongations,
the null space of $C_{i'\mu}^i$ is spanned by modes associated with both floppy modes and global rigid-body motions of which there are in total,
\begin{equation}
\label{n0}
N_0=\mathrm{nullity}(C_{i'\mu}^i).
\end{equation}

Conversely, we may consider the resulting force on each site as the linear response to a tension vector $F_i^\mu\equiv-\frac{\partial r^{i'}}{\partial r_\mu^{i}}T_{i'}$. We then obtain the equilibrium matrix,
\begin{equation}
Q_{i}^{i'\mu}T_{i'}=-F_i^\mu.
\end{equation}
Comparing with Eq.~(\ref{compatibilitymatrix}), we see that, in matrix form, $Q_i^{i'\mu}$ is the transpose of $C_{i'\mu}^i$.

In certain networks, the bonds may be placed under tension or compression while maintaining zero net force on each site, i.e., $Q_{i}^{i'\mu} T_{i'}^S=0$. Such a tensional state $T_{i'}^S$ is referred to as a SSS and is contained in the null space of $Q_{i}^{i'\mu}$. The number of SSSs in a system is thus given by
\begin{equation}
\label{ns}
N_S=\mathrm{nullity}(Q_{i}^{i'\mu})=\mathrm{nullity}(C_i^{i'\mu}).
\end{equation}
From the rank-nullity theorem and given $\mathrm{rank}(C^i_{i'\mu})=\mathrm{rank}(Q_{i}^{i'\mu})$, we obtain the index theorem \cite{lubensky},
\begin{equation}
\label{eq:Index_repeated}
N_0-N_S=dN-N_C.
\end{equation}

\added{Finally, we note the connection to the dynamical matrix, defined as}
\begin{equation}
\label{dm}
D^{j\mu}_{i\nu}=\frac{1}{m}C_{i}^{i'\mu} K_{i'}^{j'} C^{j}_{j'\nu}=\frac{1}{m}Q_{i}^{i'\mu} K_{i'}^{j'} Q_{j'\nu}^{j},
\end{equation}
\added{where $K_{i'}^{j'}\equiv\partial^2V(r_{i'})/\partial r_{j'}^2$ is the diagonal stiffness matrix.}

\end{document}